\documentstyle[aps,prd,eqsecnum]{revtex}

\def\A{{\cal A}}
\def\C{{\cal C}}
\def\H{{\cal E}}
\def\W{{\cal W}}
\def\K5{\kappa_5}

\newcommand{\square}{\kern1pt\vbox{\hrule height  1.2pt\hbox{\vrule
width 1.2pt\hskip 3pt
\vbox{\vskip 6pt}\hskip  3pt\vrule width 0.6pt}\hrule height
0.6pt}\kern1pt}
\begin{document}

\preprint{hep-th/0010133 v2}
\draft


\title{Cosmic vorticity on the brane}

\author{Helen A.~Bridgman, Karim A.~Malik and David Wands}
\address{Relativity and Cosmology Group,
School of Computer Science and Mathematics,\\ University of
Portsmouth, Portsmouth~PO1~2EG, United Kingdom}

\date{\today}

\maketitle


\begin{abstract}
  We study vector perturbations about four-dimensional brane-world
  cosmologies embedded in a five-dimensional vacuum bulk. Even in the
  absence of matter perturbations, vector perturbations in the bulk
  metric can support vector metric perturbations on the brane. We show
  that during de Sitter inflation on the brane vector perturbations in
  the bulk obey the same wave equation for a massless five-dimensional
  field as found for tensor perturbations.  However, we present the
  second-order effective action for vector perturbations and find no
  normalisable zero-mode in the absence of matter sources.  The
  spectrum of normalisable states is a continuum of massive modes that
  remain in the vacuum state during inflation.
\end{abstract}

\pacs{04.50.+h, 98.80.Cq \hfill PU-RCG 00/32, hep-th/0010133 v2\\
Accepted for publication in Phys.~Rev.~D}


\section{Introduction}

Recently a new approach to dimensional reduction has been proposed in
which matter fields are restricted to a hypersurface, or brane, while
gravity propagates in the higher-dimensional
bulk~\cite{Antoniadis,Polchinski,HW,Arkani-Hamed,Lukas}. In particular
Randall and Sundrum~\cite{RS2} have proposed a scenario in which
four-dimensional gravity may be recovered at low energies on a brane
embedded in five dimensional anti-de Sitter space (AdS$_5$).
To test this scenario we must find distinct experimental signatures of
such a model and in this paper we take up this challenge in the
context of inhomogeneous perturbations about a homogeneous and
isotropic cosmology on the brane. 

We shall consider linear perturbations about a four-dimensional
Friedmann-Robertson-Walker (FRW) spacetime embedded in a
five-dimensional vacuum bulk. Matter
perturbations couple not only to metric perturbations on the brane,
but also to perturbations in the bulk. The gauge-invariant formalism to
describe metric perturbations in brane-world models is still being
developed~\cite{Mukohyama,Langlois,Kodama,Carsten} (see
also~\cite{RoyI} for a covariant approach). As in the more familiar
four-dimensional cosmological context, metric perturbations can be
decomposed into scalar, vector and tensor perturbations with respect to
their properties on maximally symmetric spacelike
3-surfaces~\cite{Bardeen,KS,MFB92} which are defined throughout the
bulk. Solutions to the metric perturbation equations have been
presented for tensor modes on a de Sitter brane, or in the
long-wavelength limit~\cite{LMW}.  Solutions have also been found for
some scalar modes, specifically long-wavelength curvature
perturbations defined with respect to worldlines comoving with the
matter~\cite{WMLL,RoyII}, and at low-energy~\cite{Koyama}.  Unlike
tensor modes, vector modes can couple to linear matter perturbations
on the brane, but compared with scalar modes which also couple to
matter, present us with a smaller number of equations to deal with.

In this paper we will focus on vector perturbations, which are
solenoidal 3-vectors, describing cosmic
vorticity~\cite{Bardeen,KS,MFB92}. Even in the absence of matter on
the brane, they offer a much richer phenomenology in a brane-world
than is possible in ordinary four-dimensional general relativity where
the gauge-invariant vector perturbations are constrained to vanish in
the absence of any matter vorticity.  In a 4D brane-world the metric
vorticity on the brane need not vanish as it may be supported by
vorticity in the bulk gravitational field.  Vector metric
perturbations in the bulk are part of a 4D gravi-photon obtained from
dimensional reduction of five-dimensional gravity~\cite{Myung}.
We give the second-order perturbed effective action in terms of the
gauge-invariant vector perturbations in the bulk and derive their
equations of motion.

We study the particular case of de Sitter cosmology on the brane. This
is the archetypal model for slow-roll inflation in the early universe
which produces primordial spectra of perturbations on large
(super-horizon) scales at late cosmic times from an initial vacuum
state of small scale fluctuations in light fields.
We shall show that in this special case of de Sitter inflation on the
brane, the vector perturbations obey the same wave equation found for
tensor perturbations, permitting a massless zero-mode. This raises the
possibility that long-wavelength vector modes, entirely absent in
conventional models of inflation, could be generated from vacuum
fluctuations, just as happens for tensor modes~\cite{LMW}.  However,
we will show that these effectively massless modes are
non-normalisable in the bulk, corresponding to a divergent action and
hence are absent in the spectrum of quantum fluctuations.
This is complements results previously obtained for bulk vector fields
with a Minkowski brane embedded in AdS$_5$~\cite{bulkvectors}.

\section{Vector perturbations}

\subsection{Background model}

We will assume that the gravitational field in the bulk obeys the
five-dimensional vacuum Einstein equations
\begin{equation}
\label{5DEinstein}
{}^{(5)}G_{AB} = - \kappa_5^2 \, {}^{(5)}\Lambda \, {}^{(5)}g_{AB}
\,,
\end{equation}
where
$\kappa_5^2$ is the five-dimensional gravitational constant and
${}^{(5)}\Lambda$ the vacuum energy in the bulk. The gravitational
field is also subject to appropriate boundary conditions at the brane.  
The energy-momentum tensor for matter, $T_{\mu\nu}$, restricted to an
infinitesimally thin hypersurface, and the brane tension, $\lambda$,
causes a discontinuity in the extrinsic curvature, $K_{\mu\nu}$, given
by the junction conditions~\cite{Israel,BDL,SMS}
\begin{equation}
\label{junction}
\left[ K_{\mu\nu} \right]^+_-
 = - \kappa_5^2 \left( \frac{\lambda}{3} g_{\mu\nu}
+ T_{\mu\nu} - {1\over3} g_{\mu\nu} T \right) \,,
\end{equation}
where $K_{\mu\nu}=g_\mu^Ag_\nu^B\nabla_An_B$, $n^A$ is the spacelike
unit-vector normal to the brane, and the projected metric on the brane
is given by
\begin{equation}
\label{gmunu}
g_{AB}={}^{(5)}g_{AB}-n_A n_B \,.
\end{equation}
If we assume that the brane is located at a Z$_2$-symmetric orbifold
fixed point at $y=0$, then the energy-momentum tensor for matter,
$T_{\mu\nu}$, and the brane tension, $\lambda$, determine the
extrinsic curvature close to the brane
\begin{equation}
\label{Kmunu}
K_{\mu\nu} = -\frac{\kappa_5^2}{2} \left( \frac{\lambda}{3} g_{\mu\nu}
+ T_{\mu\nu} - \frac{1}{3} T g_{\mu\nu} \right) \,.
\end{equation}

In order to study inhomogeneous perturbations we will pick a specific
form for the unperturbed 5-d spacetime metric that accommodates
spatially flat FRW cosmological solutions on the brane (at any
constant-$y$ hypersurface),
\begin{equation}
\label{backmetric}
ds_5^2 = - n^2(t,y) dt^2 + a^2(t,y) \delta_{ij}dx^i dx^j
 + dy^2 \,,
\end{equation}
which includes anti-de Sitter spacetime as a special case.
Cosmological solutions of this form have been extensively
studied in the literature~\cite{bent,BDL,etc,BDEL,Flanagan}.

\subsection{Metric and matter perturbations}

We consider arbitrary linear vector perturbations, $V^i$, about the
background metric given in Eq.~(\ref{backmetric}) that are solenoidal
3-vectors on spatial hypersurfaces of constant $t$ and $y$, i.e,
$\partial_i V^i=0$. The reason for splitting generic metric
perturbations into scalar, vector and tensor modes is that they are
decoupled in the first-order perturbation equations and hence their
solutions can be derived independently of one another.

We can write the most general vector metric perturbation to
first-order as 
\begin{equation}
\label{pertmetric}
{}^{(5)}g_{AB}= \left(
\begin{array}{ccc}
-n^2 & -a^2S_i & 0 \nonumber\\
-a^2S_j &
a^2\left[ \delta_{ij} + F_{i|j} + F_{j|i}
\right] & -a^2S_{yi}
 \nonumber\\
0 & -a^2S_{yi} & 1 \nonumber
\end{array}
\right) \,,
\end{equation}
where $S_i$, $F_i$, and $S_{yi}$ are solenoidal
(divergence-free) 3-vectors on hypersurfaces of constant $t$ and $y$.
In a four-dimensional FRW metric we have only $S_i$ and
$F_i$~\cite{Bardeen,KS,MFB92}.

We will also decompose the vector perturbations into Fourier modes on
the 3-space, with time and bulk dependent amplitudes
\begin{eqnarray}
S_i &=& S(t,y) \hat{e}_i(x) \,,\\
S_{yi} &=& S_y(t,y) \hat{e}_i(x) \,,\\
F_i &=& F(t,y) \hat{e}_i(x) \,,
\end{eqnarray}
where $\hat{e}_i(x)$ is a solenoidal ($\partial^i\hat{e}_i=0$) unit
eigenvector of the spatial Laplacian such that
$\partial^j\partial_j\hat{e}_i=-k^2\hat{e}_i$.

Under an arbitrary vector gauge transformation~\cite{Bardeen}
$x^i\to x^i+\delta x^i$ (where $\partial^i\delta x_i=0$) we find
\begin{eqnarray}
S_i &\to& S_i + \dot{\delta x}_i \,,\\
S_{yi} &\to& S_{yi} + {\delta x}_i' \,,\\
F_i &\to& F_i - \delta x_i \,.
\end{eqnarray}
Thus there are essentially only two gauge-invariant variables in the
bulk which can be written as
\begin{eqnarray}
\tau &\equiv& S + \dot{F} \,,\\
\sigma &\equiv& S_y + F' \,,
\end{eqnarray}
though we will also find it useful to introduce a third
gauge-invariant combination
\begin{equation}
\label{defDelta}
\Delta \equiv S' - \dot{S}_y = \tau' - \dot\sigma \,.
\end{equation}
In four-dimensional general relativity there is only one
gauge-invariant vector metric perturbation, $\tau$~\cite{KS}.

The perturbed energy-momentum tensor for matter on the brane can be
given as
\begin{equation}
\label{Tmunu}
T^\mu_\nu = 
\left(
\begin{array}{cc}
-\rho & (\rho+P) \left( v_j - S_j|_{y=0} \right) \\
-(\rho+P)v^i & P\delta^i_j + \delta\pi^i_j
\end{array}
\right) \,.
\end{equation}
Under a 4D gauge transformation on the brane, $x^i\to x^i+\delta
x^i|_{y=0}$, the velocity perturbation transforms as $v^i\to v^i +
\dot{\delta x}{}^i|_{y=0}$ so that both the momentum and anisotropic
pressure perturbations,
\begin{eqnarray}
\label{defq}
(\rho + P) (v_i - S_i|_{y=0}) &\equiv& \hat{e}_i \, \delta p(t) \,,\\
\label{defpi}
\delta\pi^i_j &\equiv& (\partial^i\hat{e}_j + \partial_j\hat{e}^i)
\, \delta\pi(t) \,,
\end{eqnarray}
are gauge-invariant.

\subsection{Equations of motion}

The second-order perturbed Einstein-Hilbert action for gravity in the
bulk yields the effective action for the metric perturbations
\begin{equation}
\label{effaction}
\delta S = \int dt\,dy\, \left\{ - {1\over2} {a^5\over n} \Delta^2
 + {1\over2}na^3k^2\sigma^2 - {1\over2}{a^3\over n}k^2 \tau^2 \right\} \,.
\end{equation}
The three equations of motion can be derived either from first-order
perturbations of the five-dimensional Einstein
equation~(\ref{5DEinstein})~\cite{Mukohyama,Kodama,Langlois}, or by
extremising $\delta S$ with respect to variations in $F$, $S$, and
$S_y$. In either case we obtain
\begin{eqnarray}
\label{dottau}
{1\over n^2} \left\{ \dot\tau + \left(3{\dot{a}\over a} -
{\dot{n}\over n}\right) \tau \right\} &=& \sigma' + \left(3{a'\over a} +
{n'\over n} \right) \sigma \,,\\
\label{deltaprime}
{k^2\over a^2} \tau &=& \Delta
' + \left( 5{a'\over a} - {n'\over n} \right)
\Delta \,,\\
\label{dotdelta}
{k^2 n^2\over a^2} \sigma &=& \dot\Delta + \left( 5{\dot{a}\over a} -
{\dot{n} \over n} \right) \Delta \,.
\end{eqnarray}

{}From these we can obtain coupled second-order evolution
equations for $\tau$ or $\sigma$:
\begin{eqnarray}
\label{taueom}
{1\over n^2} \left\{ \ddot\tau + 3\left( {\dot{a}\over a} -
{\dot{n}\over n} \right) \dot\tau
 + \left( 3{\ddot{a}\over a} - 3{\dot{a}^2\over a^2} -
6{\dot{a}\dot{n}\over an} - {\ddot{n}\over n} + 3{\dot{n}^2\over n^2}
\right) \tau \right\} + {k^2\over a^2}\tau \nonumber\\
=
\tau'' + \left(5{a'\over a} - {n'\over n} \right) \tau'
+ 2\left( {n'\over n} - {a'\over a} \right) \dot\sigma
+ \left( 3{\dot{a}\over a} + {\dot{n}\over n} \right)^\prime \sigma \,,\\
\label{rhoeom}
{1\over n^2} \left\{ \ddot\sigma + \left( 5{\dot{a}\over a} -
{\dot{n}\over n} \right) \dot\sigma - 2{\dot{a}\over a} \tau'
 + \left( 3{\dot{a}'\over a} -3{\dot{a}a'\over a^2} - {\dot{n}'\over
n} + {\dot{n}n' \over n^2} \right) \tau \right\} + {k^2 \over a^2}\sigma
\nonumber \\
=
\sigma''+\left( 3{a'\over a} - {n'\over n} \right) \sigma' + \left(
3{a''\over a} - 3{a^{\prime2} \over a^2} + {n''\over n} +
{n^{\prime2}\over n} + 6{n'a'\over na} \right) \sigma \,,
\end{eqnarray}
and a decoupled evolution equation for $\Delta$:
\begin{eqnarray}
\label{deltaeom}
{1\over n^2} \left\{ \ddot\Delta + \left( 7{\dot{a}\over a}
-3{\dot{n}\over n} \right) \dot\Delta \right\} - \left\{ \Delta'' +
\left( 7{a'\over a} - {n'\over n} \right) \Delta' \right\} + {k^2\over
a^2} \Delta
\nonumber \\
=
 \left\{ 5{a''\over a} + 5{a^{\prime2}\over a^2} -2{a'n'\over an} - {n''\over
n} + {n^{\prime2} \over n^2} - {1\over n^2} \left( 5{\ddot{a}\over a}
+ 5 {\dot{a}^2 \over a^2} -12 {\dot{a}\dot{n} \over an} -
{\ddot{n}\over n} +3{\dot{n}^2\over n^2} \right) \right\} \Delta \,.
\end{eqnarray}

\subsection{Master variable}
\label{SSectmaster}

The form of the constraint equation~(\ref{dottau}) implies that
there exists a function $\Omega$ such that
\begin{equation}
\label{tauOmega}
\tau = {n\over a^3}\Omega' \,, \qquad
\label{sigmaOmega}
\sigma = {1\over na^3}\dot\Omega \,.
\end{equation}
Substituting this into the evolution equations~(\ref{deltaprime})
and~(\ref{dotdelta}) they can be integrated to yield
\begin{equation}
\label{DeltaOmega}
\Delta = {n\over a^5} k^2 \Omega \,,
\end{equation}
(where an arbitrary constant of integration has been absorbed into the
definition of $\Omega$) and hence we obtain a single {\em master
equation}~\cite{Mukohyama,Kodama}
\begin{equation}
\label{Omegaeom}
{1\over n^2} \left[ \ddot\Omega - \left( 3{\dot{a}\over a} +
{\dot{n}\over n} \right) \dot\Omega \right] - \left[ \Omega'' - \left(
3{a'\over a} - {n'\over n} \right) \Omega' \right] + {k^2\over a^2}
\Omega = 0 \,.
\end{equation}

Substituting Eqs.~(\ref{tauOmega}), (\ref{sigmaOmega})
and(\ref{DeltaOmega}) into Eq.~(\ref{effaction}),
the effective action for perturbations in the bulk becomes
\begin{equation}
\label{Omegaeffaction}
\delta S = k^2 \int dt dy {n\over 2a^3} \left[ {1\over n^2}\dot\Omega^2
  - \Omega^{\prime2} - k^2 \Omega^2 \right] \,,
\end{equation}
whose variation with respect to $\Omega$ gives the equation of
motion~(\ref{Omegaeom}).

\subsection{Junction conditions}

The first-order perturbations in the extrinsic curvature of constant-$y$
hypersurfaces due to vector perturbations are
\begin{eqnarray}
\delta K^0_i &=& {a^2\over2n^2} \Delta \, \hat{e}_i \,,\\
\delta K^i_j &=& {1\over2} \sigma \left( \partial^i\hat{e}_j +
  \partial_j\hat{e}^i \right) \,.
\end{eqnarray}
Equations~(\ref{Kmunu}), (\ref{Tmunu}), (\ref{defq}) and~(\ref{defpi})
thus yield
\begin{eqnarray}
\label{matchq}
\Delta_b &=& - \kappa_5^2\, {n^2\over a^2}\, \delta p \,, \\
\label{matchpi}
\sigma_b &=& - \kappa_5^2 \, \delta\pi \,,
\end{eqnarray}
where the subscript ``$b$'' denotes quantities evaluated on the
brane. 
Thus we find that the gauge-invariant metric perturbation, $\sigma$,
vanishes on the brane for a perfect fluid with vanishing anisotropic
pressure. 
$\Delta$, defined in Eq.~(\ref{defDelta}), is constrained to vanish on
the brane in the absence of any solenoidal 3-momentum. 
However the gauge-invariant vector perturbation, $\tau$, remains
non-vanishing in general on the brane, even when the matter on the
brane possesses no vorticity.  
Its first derivative normal to the brane, $\tau'_b$, is zero if
both the momentum and anisotropic pressure perturbations vanish, but
$\sigma'$ may still be non-zero at the brane and the evolution of
$\tau$ on the brane cannot in general be decoupled from the evolution
of $\tau$ and $\sigma$ in the bulk.

In terms of the master variable $\Omega$ we have
\begin{eqnarray}
\label{matchqOmega}
k^2 \Omega_b &=& - \kappa_5^2\, n_b a_b^3\, \delta p \,, \\
\label{matchpiOmega}
\dot\Omega_b &=& - \kappa_5^2\, n_b a_b^3\, \delta\pi \,.
\end{eqnarray}
Taken together these are consistent with usual momentum conservation
equation for matter on the brane:
\begin{equation}
\label{mtmconservation}
\dot{\delta p} + \left( 3{\dot{a}_b\over a_b} + {\dot{n}_b\over n_b} \right)
\delta p = k^2 \delta\pi \,.
\end{equation}

\section{de Sitter inflation on the brane}

The only separable solution for the background metric
\begin{equation}
n=\A(y) \,, \qquad a=a_b(t) \A(y) \,,
\end{equation}
describes de Sitter expansion on the brane
($\dot{a}_b/a_b=H=$constant) in an AdS$_5$ bulk.
Solving for the metric in the bulk yields~\cite{bent,LMW}
\begin{equation}
\label{Ay}
\A(y) = {H\over\mu} \sinh \left[ \mu \left( y_h - |y| \right) \right] \,,
\end{equation}
where $\mu$ is the anti-de Sitter curvature scale, 
and there is a Cauchy horizon at $|y|=y_h$ where $\A(y)=0$~\cite{bent}.

\subsection{Metric perturbation, $\tau$}

In this case the equation of motion~(\ref{taueom}) for $\tau$
decouples from $\sigma$ in the bulk and we have the equation of motion
\begin{equation}
\label{deSittertaueom}
\ddot{\tau} +3H\dot\tau + {k^2\over a_b^2} \tau = \A^2 \tau'' +
4\A\A' \tau' \,.
\end{equation}
This has exactly the same form as the equation for tensor
perturbations with de Sitter expansion on the brane. Following
Refs.~\cite{GS,LMW} we separate
\begin{equation}
\tau(t,y) = \int dm \, \varphi_m (t) \H_m (y) \,,
\end{equation}
and obtain two ordinary differential equations
\begin{eqnarray}
\ddot{\varphi}_m +3H\dot{\varphi}_m+\left[ m^2+{k^2\over
a_b^2}\right] \varphi_m &=&0\,, \label{varphieom}\\
\H_m''+4{\A'\over\A}\H_m'+ {m^2\over \A^2}\H_m &=& 0\,.
\label{Heom}
\end{eqnarray}
Each bulk eigenmode, characterised by the eigenvalue $-m^2$, becomes a
4D effective field with mass $m^2$. The general solution for the
time-dependence of a massive field in de Sitter is given by~\cite{BD}
\begin{equation}
\varphi_m(t)
 = \left( \frac{k}{a_bH} \right)^{3/2} \,
 B_\nu\left(\frac{k}{a_bH}\right) \,,
\end{equation}
where $B_\nu$ is a linear combination of Bessel functions of order
$\nu=\sqrt{(9/4)-(m^2/H^2)}$. Fields with $m\leq3H/2$ are light during
inflation and vacuum fluctuations on small scales ($k\gg a_bH$) can
give rise to a spectrum of perturbations on large scales ($k\ll a_bH$)
at late times. By contrast heavy modes (with $m>3H/2$) remain in the
vacuum state with essentially no fluctuations on large scales.

In Ref.~\cite{LMW} the amplitude of vacuum fluctuations for tensor
modes was calculated by finding the spectrum of modes with finite
perturbed effective action in the bulk. In that case a discrete
massless ($m=0$) mode and a continuum of massive modes ($m>3H/2$) was
found, so that during inflation the massive modes remain in their
vacuum state and only the massless mode acquires a spectrum of
fluctuations on large scales~\cite{LMW}.

For $m=0$ our bulk eigenmode has the general solution
\begin{equation}
\label{Hzeromode}
\H_0(y) = C_1 + C_2 \int {dy \over {\cal A}^4} \,.
\end{equation}
Although this is divergent for $C_2\neq0$ at the horizon where ${\cal
  A}\to0$, the boundary conditions at the brane, given in
Eqs.~(\ref{matchq}) and~(\ref{matchpi}) require $\tau'_b$ and hence
$C_2=0$ in the absence of solenoidal momentum and anisotropic pressure
on the brane ($\delta p = 0$, $\delta\pi=0$), as in the case of
slow-roll inflation driven by scalar fields.

In the case of tensor perturbations studied in Ref.~\cite{LMW} this
was sufficient to keep the perturbed effective action finite, leading
to the generation of large-scale tensor perturbations in this massless
mode from small-scale vacuum fluctuations during inflation on the
brane. However in the present case of vector perturbations, the
effective action~(\ref{effaction}) includes contributions from the
other gauge-invariant metric perturbation, $\sigma$. To evaluate the
corresponding solution for $\sigma$ we turn to the master variable
$\Omega$ introduced in subsection~\ref{SSectmaster}.

\subsection{Master variable}

For a de Sitter brane we are able to separate
\begin{equation}
\Omega(t,y) = \int \, dm \, \omega_m(t) \, \W_m(y) \,,
\end{equation}
where the equation of motion~(\ref{Omegaeom}) requires
\begin{eqnarray}
\label{omegaeom}
\ddot\omega_m - 3H \dot\omega_m + \left[ m^2 + {k^2\over a_b^2}
\right] \omega_m &=& 0 \,,\\
\label{Weom}
\W_m'' - 2{\A'\over\A} \W_m' + {m^2\over\A^2}\W_m &=& 0 \,.
\end{eqnarray}
{}From the definition of $\tau$ in Eq.~(\ref{tauOmega}) we see that
$\omega_m=a_b^3\varphi_m$, which is consistent with
Eqs.~(\ref{varphieom}) and~(\ref{omegaeom}).

Following the original paper of Randall and Sundrum~\cite{RS2} for a
Minkowski brane, and the analysis of tensor modes for a de Sitter
brane~\cite{GS,LMW}, we define
\begin{equation}
\Psi_m = \A^{-3/2} \W_m \,, \qquad
dz = {dy \over \A} \,,
\end{equation}
and hence $z\to\infty$ as $y\to y_h$.
The canonical field $\Psi_m$ then obeys the Schr\"{o}dinger-like
equation 
\begin{equation}
{d^2 \Psi_m \over dz^2} - V(z)\Psi_m = -m^2 \Psi_m \,,
\end{equation}
with the effective potential
\begin{equation}
V(z) = {3\over4}\mu^2\A^2 + {9\over4}H^2 + 3H\delta(z) \,.
\end{equation}
Like the original `Volcano potential' of Ref.~\cite{RS2}, $V(z)$
decreases as the `warp factor' $\A(y)$ decreases away from the brane,
and like its cosmological generalisation for tensor
modes~\cite{GS,LMW} it approaches a constant value, $V\to 9H^2/4$, as
$z\to\infty$. However a crucial difference is that the Dirac
$\delta$-function, which has a negative coefficient and yields the 4D
graviton localised on the brane in Refs.~\cite{RS2,GS,LMW}, here has a
positive coefficient so that there is no bound zero-mode for vector
perturbations.

The general solution for the zero-mode, $m=0$, is given by
\begin{equation}
\label{zeroW}
\W_0(y) = \tilde{C}_2 + C_1 \int_0^y \A^2(y') dy' \,.
\end{equation}
Note that substituting this expression into Eq.~(\ref{tauOmega}) in
order to determine $\H_0(y)$ yields only the homogeneous solution,
$\H_0=C_1$ in Eq.~(\ref{Hzeromode}). 

The junction condition~(\ref{matchqOmega}) at the brane yields 
\begin{equation}
\label{matchdSOmega}
\Omega_b = a_b^3 \varphi_0 \tilde{C}_2
 = - {\kappa_5^2\over k^2} a_b^3 \delta p \,.
\end{equation}

In order to calculate the spectrum of vacuum fluctuations on the brane
we need to evaluate the effective action for each bulk eigenmode. For
each mode $m$ we obtain from Eqs.~(\ref{Omegaeffaction})
and~(\ref{Weom})
\begin{equation}
\delta S_m = \C_m^2 \int dt \, {k^2 a_b^3\over 2} \left[
  \dot\varphi_m^2 - \left( {k^2\over a_b^2} + m^2 \right) \varphi_m^2
  \right] \,,
\end{equation}
where the normalisation constant for each mode is given by
\begin{equation}
\label{Cm}
\C_m^2 = \int_{-\infty}^\infty \, |\Psi_m|^2 \, dz \,.
\end{equation}
This yields an effective action which has the standard form for a
four-dimensional field, $\bar\phi_m=k\C_m\varphi_m$, with mass $m$ in
a FRW spacetime with scale factor $a_b$, for any normalisable modes,
i.e., modes with finite $\C_m^2$.

For modes with $m^2>9H^2/4$ their asymptotic solutions for $\Psi_m$ at
large $z\gg H^{-1}$ are plane waves $\Psi\propto e^{\pm i\tilde{k}z}$,
where $\tilde{k}^2=m^2-9H^2/4$, and these modes are thus
normalisable~\cite{RS2}. 

In the absence of solenoidal momentum or pressure perturbations, the
junction conditions at the brane in Eqs.~(\ref{matchqOmega})
and~(\ref{matchpiOmega}) yields $\tilde{C}_2=0$, while allowing
$C_1\neq0$ in Eq.~(\ref{zeroW}). Although this leaves $\W_0$
finite, we have $\Psi_0 = \W_0/\A^{3/2}$, and this leads to a
divergent integral in Eq.~(\ref{Cm}) for $\C_0$. Thus {\em there is no
normalisable zero-mode in the absence of vector matter perturbations
on the brane}.

Thus, even though the gauge-invariant metric perturbation, $\tau$,
obeys the same wave equation~(\ref{deSittertaueom}) for a massless 5D
field as previously found for tensor perturbations~\cite{GS,LMW}, we
find no `light' normalisable vector modes ($m\leq3H/2$) that could
be excited by de Sitter expansion on the brane in order to generate
large scale (super-horizon) vector metric perturbations on the brane
from an initial vacuum state on small scales.

We can only construct a normalisable zero-mode if $\W_0$ vanishes at
the Cauchy horizon, corresponding to
$\tilde{C}_2=-C_1\int^{y_h}_0\A^2dy$. In this case the junction
condition~(\ref{matchdSOmega}) then requires that the momentum of
matter on the brane is $(k^2/\kappa_5^2) \tau_{b}$, which appears
to have no physical motivation.
Similarly, if we were to introduce a `regulator brane' just within the
Cauchy horizon in order to keep the normalisation constant, $\C_m^2$,
finite with $\tilde{C}_2=0$, the regulator brane would have to
possess just the right matter perturbation in order to satisfy the
junction condition. This is in contrast to the original
Randall-Sundrum zero-mode~\cite{RS2} where the regulator brane has
constant brane tension, and the normalisation of the zero-mode has a
well-defined limit as the regulator brane approaches the Cauchy
horizon. 

\section{The view from the brane}

Shiromizu, Maeda and Sasaki \cite{SMS} showed that the effective
four-dimensional Einstein equations on the brane can be obtained by
projecting the five-dimensional variables onto the 4D brane world.  If
our 4-dimensional world is described by a domain wall (3-brane)
$(M,g_{\mu\nu})$ in five-dimensional spacetime $({\cal
  M},{}^{(5)}g_{AB})$, where the induced metric, $g_{\mu\nu}$, was
defined in Eq.~(\ref{gmunu}), then, using the Gauss equation
\cite{SMS}, one obtains the 4-dimensional effective equations as
\begin{equation}
{}^{(4)}G_{\mu\nu}
 = - {1\over 2} \kappa_5^2 {}^{(5)}\Lambda
+ KK_{\mu\nu}
-K^{~\sigma}_{\mu}K_{\nu\sigma} -{1 \over 2}g_{\mu\nu}
  \left(K^2-K^{\alpha\beta}K_{\alpha\beta}\right) - E_{\mu\nu},
\label{4dEinstein}
\end{equation}
where $K_{\mu\nu}$ is the extrinsic curvature of the brane,
$K=K^\mu_{~\mu}$ is its trace, and the effect of the non-local bulk
gravitational field is described by the projected five-dimensional
Weyl tensor
\begin{equation}
E_{\mu\nu} \equiv {}^{(5)}C^E_{~A F B}n_E n^F
g_\mu^{~A} g_\nu^{~B} \,.
\label{Edef}
\end{equation}
Using the junction conditions given in Eq.~(\ref{Kmunu}), we can
give the extrinsic curvature in terms of the energy-momentum tensor on
the brane and we obtain
\begin{eqnarray}
\label{modEinstein}
{}^{(4)}G_{\mu\nu}=-{}^{(4)}\Lambda g_{\mu\nu}
+ \kappa_4^2 T_{\mu\nu}+\kappa_5^4\,\Pi_{\mu\nu}
-E_{\mu\nu}\,, \label{eq:effective}
\end{eqnarray}
%
where
\begin{eqnarray}
{}^{(4)}\Lambda &=&\frac{\kappa_5^2}{2}
\left[{}^{(5)}\Lambda +\frac{1}{6}\kappa_5^2\,\lambda^2 \right] \,,
\label{Lamda4}\\
\kappa_4^2\ &=& 8\pi G_N={\kappa_5^4\,\over 6}\lambda\,,
\label{GNdef}\\
\Pi_{\mu\nu}&=&
-\frac{1}{4} T_{\mu\alpha}T_\nu^{~\alpha}
+\frac{1}{12}TT_{\mu\nu}
+\frac{1}{8}g_{\mu\nu}T_{\alpha\beta}T^{\alpha\beta}-\frac{1}{24}
g_{\mu\nu}T^2\,.
\label{pidef}
\end{eqnarray}
The power of this approach is that the above form of the
four-dimensional effective equations of motion is independent of the
evolution of the bulk spacetime, being given entirely in terms of
quantities defined at the brane. Thus these equations apply to
brane-world scenarios with infinite or finite bulk, stabilised or
evolving.

The perturbed 4D Einstein tensor, derived for the perturbed metric
induced on the brane, is
\begin{eqnarray}
\label{delta4G0i}
\delta^{(4)}G^0_i &=& {1\over2n_b^2} \, k^2\tau_b \, \hat{e}_i \,,\\
\label{delta4Gij}
\delta^{(4)}G^i_j &=& {1\over2n_b^2}\left[ 
 \dot\tau_b
 + \left( 3{\dot{a}_b\over a_b} - {\dot{n}_b\over n_b} \right) \tau_b
 \right] \left(
  \partial^i\hat{e}_j + \partial_j\hat{e}^i \right) \,.
\end{eqnarray}
Note that only $\tau_b$ appears in these expressions as it is the
only gauge-invariant vector metric perturbation of the 4D
metric~\cite{KS}. 

Substituting the perturbed energy-momentum tensor given in
Eq.~(\ref{Tmunu}) into Eq.~(\ref{modEinstein}) gives the contribution
of the matter on the brane
\begin{eqnarray}
\label{delta4G0i1}
\kappa_4^2 \delta T^0_i + \kappa_5^4\,\delta\Pi^0_i
 &=& \kappa_4^2 \left( 1 + {\rho\over\lambda}
\right) \delta p \, \hat{e}_i \,,\\
\label{delta4Gij1}
\kappa_4^2 \delta T^i_j + \kappa_5^4\,\delta\Pi^i_j
 &=& \kappa_4^2 \left( 1 - {\rho+3P\over2\lambda}
\right) \delta\pi^i_j \,.
\end{eqnarray}
where the contribution from $\delta\Pi^\mu_\nu$ becomes negligible for
$\rho\ll\lambda$ and $P\ll\lambda$.  The junction condition at the
brane, given in Eq.~(\ref{junction}), relates these matter
perturbations to metric perturbations at the brane:
\begin{eqnarray}
\kappa_4^2 \left( 1 + {\rho\over\lambda}
\right) \delta p \, \hat{e}_i 
 &=& a_b^2 \left( {a_b'\over a_b} \Delta_b \right)
\hat{e}_i 
 \,,\\ 
\kappa_4^2 \left( 1 - {\rho+3P\over2\lambda}
\right) \delta\pi^i_j &=& {1\over2} \left( {n_b'\over n_b} + {a_b'\over a_b}
\right) \sigma_b 
\left( \partial^i\hat{e}_j + \partial_j\hat{e}^i \right)
 \,. 
\end{eqnarray}

Finally the contribution of metric perturbations in the bulk to the
modified Einstein equations on the brane is given by the projected
Weyl tensor $\delta E^\mu_\nu$. For vector perturbations, in terms of
our gauge invariant variables, we have
\begin{eqnarray}
\label{eq:deltaE0i} \delta {E^0_i} &=&  - \frac{1}{6 n_b^2}
\left\{2a_b^2 \left[ {\Delta}_b' + \left( 2 \frac{a_b'}{a_b} -
\frac{n_b'}{n_b} \right) {\Delta}_b \right] + k^2 {\tau}_b\right\}
 \hat{e}_i
\,,\\
\label{eq:deltaEij} 
\delta {E^i_j} &=& - \frac{1}{6 n_b^2} \left\{
\dot{\tau}_b + \left( 3\frac{\dot{a}_b}{a_b} - \frac{\dot{n}_b}{n_b} \right)
 \tau_b
 + n_b^2 \left[ 2 \sigma_b' + \left( 3\frac{a_b'}{a_b} -
     \frac{n_b'}{n_b} \right) 
\sigma_b \right] \right\}
\left( \partial^i\hat{e}_j + \partial_j\hat{e}^i \right)
\, .
\end{eqnarray}
Substituting Eqs.~(\ref{delta4G0i})--(\ref{eq:deltaEij}), into the
four-dimensional Einstein equations~(\ref{eq:effective}), yields the
first two five-dimensional field equations~(\ref{dottau})
and~(\ref{deltaprime}).  The remaining five-dimensional field
equation~(\ref{dotdelta}) can be obtained directly from the
conservation of the matter momentum~(\ref{mtmconservation}), on the
brane using Eqs.~(\ref{matchq}) and~(\ref{matchpi}).  However the
equations at the brane do not in general yield a closed set of
equations on the brane. Although $\sigma_b$ and $\Delta_b$ are
determined by the junction conditions~(\ref{matchq})
and~(\ref{matchpi}), the behaviour of their y-derivatives, $\sigma_b'$
and $\Delta_b'$, must be determined by solving the five-dimensional
equations in the bulk.

The projected Weyl tensor acts as an effective energy-momentum tensor
on the brane, whose effective momentum can be written, using
Eqs.~(\ref{eq:deltaE0i}) and (\ref{deltaprime}), as
\begin{equation}
\kappa_4^2 \widetilde{\delta p}
= {a_b^2\over 2n_b^2} \left\{ \Delta_b' + \left( 3{a_b'\over a_b} -
  {n_b'\over n_b} 
  \right) \Delta_b \right\} \,,
\end{equation}
and the effective anisotropic pressure can be written, using 
Eqs.~(\ref{eq:deltaEij}) and (\ref{dottau}), as
\begin{equation}
\kappa_4^2 \widetilde{\delta \pi} 
= {1\over2} \left\{ \sigma_b' + 2{a_b'\over a_b} \sigma_b \right\} \,.
\end{equation}
Even when there are no vector perturbations in the matter
energy-momentum tensor on the brane, the projected 5D Weyl tensor can
supply an effective anisotropic momentum and stress and hence support
vector perturbations in the induced 4D metric.

The contracted Bianchi identities ($\nabla_\mu{}^{(4)}G^\mu_\nu=0$)
and energy-momentum conservation for matter on the brane ($\nabla_\mu
T^\mu_\nu=0$) ensures, from Eq.~(\ref{modEinstein}), that $\nabla_\mu
E^\mu_\nu=\kappa_5^4\,\nabla_\mu\Pi^\mu_\nu$.  The interaction with
the quadratic energy-momentum tensor, $\Pi^\mu_\nu$, gives rise to the
momentum conservation equation for the Weyl-fluid
\begin{equation}
\label{deltapeom}
\dot{\widetilde{\delta p}} + \left( 3{\dot{a}_b \over a_b}
  + {\dot{n}_b \over n_b} \right) \widetilde{\delta p}
 = k^2 \widetilde{\delta \pi}
+ 6 \left( {\rho+P\over\lambda} \right) \left( {\dot{a}_b \over a_b} \delta p
 - {k^2 \over2}\delta\pi \right) \,.
\end{equation}
Thus the Weyl-momentum is conserved in the absence of vorticity
in the matter, or when $(\rho+P)/\lambda$ is negligible.

\section{Conclusions}

We have studied the nature of vector perturbations of brane-world
cosmologies embedded in a five-dimensional bulk described by vacuum
Einstein gravity.  By a simple extension of the standard
four-dimensional cosmological studies~\cite{Bardeen,KS,MFB92}, vector
perturbations are described by divergence-free 3-vectors on Euclidean
spatial hypersurfaces of fixed time $t$ and bulk coordinate $y$ which
foliate the five-dimensional spacetime~\cite{Mukohyama,Kodama}.

There is only one gauge-invariant vector perturbation, $\tau$, of the
four-dimensional FRW metric induced on the brane. However in the
five-dimensional bulk we can define a second gauge-invariant vector
metric perturbation, $\sigma$, and in general the evolution of the
vector perturbation on the brane cannot be decoupled from the bulk
vector perturbation leading to coupled equations of motion.
Even in the case of vanishing matter perturbations, the vector metric
perturbations in the bulk can support vector metric perturbations on
the brane, in contrast to four-dimensional Einstein gravity where the
vector metric perturbations are constrained to vanish in the limit of
vanishing matter vorticity~\cite{Bardeen,KS}. 

In the case of de Sitter expansion on the brane $\tau$ decouples from
$\sigma$ and obeys the equation of motion for a massless
five-dimensional field, as previously found for tensor
perturbations~\cite{GS,LMW}. In the case of tensor perturbations this
leads to a spectrum of large-scale (super-horizon) perturbations being
generated from an initial quantum vacuum state on sub-horizon
scales~\cite{LMW}. The prediction of a spectrum of vector
perturbations from cosmological inflation on the brane would be a
distinctive observational prediction of the brane-world cosmology, but
we have shown that there is no normalisable zero-mode for the vector
perturbations in the bulk that respect the vacuum junction conditions
at the brane. The spectrum of normalisable modes (with finite 4D
effective action) is a continuum of modes with masses $m>3H/2$, where
$H$ is the Hubble expansion rate. This includes the case of a
Minkowski brane in the limit $H\to0$. 

The effective action for vector perturbations is most concisely
written in terms of the ``master variable'', $\Omega$, from which both
$\tau$ and $\sigma$ can be derived~\cite{Mukohyama,Kodama}. Following
the approach of Randall and Sundrum~\cite{RS2} we derive the
Schr\"{o}dinger-like equation for the canonically normalised variable
with a modified ``Volcano'' potential. The absence of a normalisable
zero-mode is due to the change in sign of the Dirac delta-function at
the brane. 
The only way to obtain a normalisable vector zero-mode seems to be to
have a matter source on the brane whose momentum cancels out this
effect.

Except in the case of a de Sitter brane, the equations of motion in
the bulk are not separable, and we cannot decompose the
five-dimensional perturbations into Kaluza-Klein modes.
The effect of the bulk metric perturbations on the brane can be
described by the projection of the five-dimensional Weyl tensor on the
brane~\cite{SMS}. This is seen by the brane-world observer as a
form of dark matter on the brane with trace-free energy-momentum
tensor which may include anisotropic stresses.
We have shown how the bulk vector perturbations would be interpreted
by an observer in the brane-world as an effective momentum and
anisotropic stress on the brane, arising from this projected bulk Weyl
`fluid'~\cite{SMS}. 
%
%
The contracted Bianchi identities on the brane yield an effective
momentum conservation equation for the Weyl fluid, but the effective
anisotropic stress cannot be determined from quantities locally on the
brane, a consequence of the five-dimensional origin of the
perturbations. 
Although the anisotropic stress decouples from the momentum evolution
on large scales, it is needed in order to determine the
gauge-invariant metric perturbation, $\tau$, which enters the
Sachs-Wolfe formula for anisotropies in the cosmic microwave
background~\cite{Ruth}. 
A similar effect was found recently for scalar metric
perturbations~\cite{scalar}.

It has been suggested~\cite{Langlois} that bulk metric perturbations
might thus provide `active seeds' for cosmological perturbations,
similar to topological defects.
Such a scenario could be realised by large-scale fluctuations of the
master variable, $\Omega$.
While spatial gradients on the brane are negligible, $\Omega=$constant
throughout the bulk is an approximate solution of the bulk equation of
motion~(\ref{Omegaeom}) which yields no vector perturbations on the
brane ($\tau$ and $\sigma$ both vanish). However, when spatial
gradients become important (e.g., after horizon entry during the
radiation or matter dominated eras) a non-zero constant $\Omega$ is no
longer a solution of the bulk equation of motion and the resulting
time and bulk variation of $\Omega$ would generate vector
perturbations on the brane.

Although such vector perturbations can appear apparently from nothing
on the brane, five-dimensional gravity is a causal system, and any
large-scale variations of $\Omega$ can be traced back to some initial
conditions in the bulk. 
Such a scenario seems possible for scalar or tensor perturbations
which possess a normalisable zero-mode which may be excited during a
period of inflation in the early universe.
However we have shown that the absence of a normalisable zero-mode for
vector perturbations in the bulk means that a period of de
Sitter inflation in the early universe leaves effectively no
large-scale vector perturbations in the bulk. Massive modes remain
in their vacuum state, and $\Omega=0$ remains a solution to the bulk
equation of motion at all subsequent times.

\acknowledgments

The authors are grateful to Fabio Finelli and Roy Maartens for useful
comments and discussions.
HAB is supported by the EPSRC, and DW by the Royal Society.
Algebraic computations of tensor components were performed using
the GRTensorII package for Maple.

\end{document}